# Fluorescence microscopy of single autofluorescent proteins for cellular biology

Laurent Cognet*, Françoise Coussen†, Daniel Choquet† & Brahim Lounis*

\* *Centre de Physique Moléculaire Optique et Hertzienne - CNRS UMR 5798 et Université Bordeaux 1, 351 Cours de la Libération, 33405 Talence, France*

† *Laboratoire de Physiologie Cellulaire de la Synapse - CNRS UMR 5091 et Université Bordeaux 2, Institut François Magendie, 1 rue Camille Saint-Saëns 33077 Bordeaux, France*

In this paper we review the applicability of autofluorescent proteins for single-molecule imaging in biology. The photophysical characteristics of several mutants of the Green Fluorescent Protein (GFP) and those of *Ds*Red are compared and critically discussed for their use in cellular biology. The alternative use of two-photon excitation at the single-molecule level or Fluorescence Correlation Spectroscopy is envisaged for the study of individual autofluorescent proteins. Single-molecule experiments performed in live cells using eGFP and preferably eYFP fusion proteins are reviewed. Finally, the first use at the single-molecule level of citrine, a more photostable variant of the eYFP is reported when fused to a receptor for neurotransmitter in live cells.

Nous présentons une revue sur l'utilisation des protéines autofluorescentes en microscopie de molécules uniques en biologie. Les caractéristiques photophysiques de plusieurs mutants de la protéine fluorescente verte (la « GFP ») ainsi que celles de la *Ds*Red y sont comparées et discutées de manière critique en vue de leur utilisation dans des cellules vivantes. Des méthodes alternatives d'excitation, telle l'excitation biphotonique, ou d'analyses, telle la méthode de spectroscopie par corrélation de fluorescence sont envisagées. Nous rendons compte d'expériences utilisant la eGFP et préférentiellement la eYFP au niveau de la molécule unique dans des cellules vivantes. Nous reportons enfin, la première utilisation au niveau de la molécule individuelle de la citrine, un mutant de la eYFP plus résistant au photoblanchiment et son application à l'étude de la dynamique de récepteurs de neurotransmetteurs individuels dans la membrane de cellules vivantes.

Mots clés : détection de molécules uniques, GFP, *Ds*Red, microscopie de fluorescence, excitation à deux photons, spectroscopie par corrélation de fluorescence.

Research in the post-genomic era is now enhanced by applications of emerging sensitive detection techniques with modern biological methods. Single-molecule spectroscopy (SMS) is one of the most exciting developments since it plays a unique role in the elucidation of the dynamics of an increasingly broad range of complex systems[1, 2]. By removing the averaging inherent to ensemble measurements, SMS yields a measure of the distribution of molecular properties which is of importance in systems which display static or time-dependent heterogeneity. Concurrently, developments in fluorescence microscopy have made the observation of single molecules of biological interest possible under a wide range of experimental conditions[3]. This versatility can be exploited to examine the molecular complexity of many biological systems.

One critical challenge to the application of SMS techniques to biomolecules, especially in live cells, is the introduction of a fluorophore that acts as a reporter of activity, local environment, or spatial location. While certain important proteins display useful amounts of visible fluorescence arising from native cofactors[4], most require the attachment of an extrinsic fluorophore to serve as a probe[5-8]. There is a large variety of labeling methods for proteins applicable to in vitro assays[9], including several new developments utilizing semiconductor quantum dots[10, 11] and highly photostable fluorophores[12]. However, for labeling in live cells, those optimized fluorescence labels are of limited use. One of the most convenient, common, and benign ways to specifically label proteins in vivo is to construct a fusion with an autofluorescent protein from the jellyfish *Aequoria victoria* or one of its variants (the 'XFPs' [13] where X stands for the dominant color of the emission spectrum: C for cyan, G for green and Y for yellow) or from the *Discosoma genus* of coral[14], *Ds*Red. This genetic fusion of naturally fluorescent proteins to a protein of interest ensures a reproducible 1:1 stoichiometry in the cell, without the need for external chemical reactions and with the hope of less interference with the biological function and vitality of the cell.

Combining single-molecule microscopy with genetic labeling by autofluorescent proteins has proved to be difficult due mainly to interference of the single-molecule fluorescence signal with background fluorescence created by other cellular constituents and due to the photophysical properties of the autofluorescent proteins themselves. Still, good knowledge of those characteristics allowed for successful visualization of single autofluorescent proteins in live cells[15-17]. In this review, we will discuss the photophysical parameters of commercially available autofluorescent proteins essential for single-molecule detection research[15]. Applications of detection of single biomolecules labeled with autofluorescent proteins will then be discussed.

## 1. Introduction

Single molecule fluorescence detection is successfully obtained when careful control of the fluorescence background is achieved. Indeed, cellular autofluorescence might hinder the weak signal of individual fluorophores. Different techniques are commonly used in conventional fluorescence microscopy to diminish cellular autofluorescence: confocal microscopy or total internal microscopy for instance[18]. Nevertheless, none of those techniques will be presented in this review since they present some disadvantages which are discussed later. Thus, we will address the issue throughout the whole review of the detection of single autofluorescent proteins in live cells using wide field fluorescence microscopy. The basic setup is presented on Fig. 1. It is composed of an inverted microscope equipped with a high NA oil-immersion objective (typ. ×100, NA=1.4). The samples are then illuminated for a few milliseconds by a single line laser (an Ar+ laser for the XFPs and a frequency doubled Nd:YAG laser for *Ds*Red) at an illumination intensity of a few kW/cm$^2$. By using an appropriate filter combination for a given fluorophore and highly sensitive CCD camera systems, the total detection efficiency of a typical experimental setup is in the range of 5-10%.

Pioneer studies at the level of individual autofluorescent proteins were performed in vitro situation with purified protein immersed in biocompatible matrices [19-22] or buffer[23]. Those studies have revealed anomalous fluorescence properties. The most striking behavior is the reversible photobleaching as revealed by Dickson and coworkers or "blinking" [24], which have remained hidden in previous bulk studies. These phenomenon can be interpreted as a consequence of the presence of different non-fluorescent states. The lifetime and transitions between these states have been revealed by Fluorescence Correlation Spectroscopy (FCS)[25, 26]. A noticeable feature concern the broad range of the lifetime of these dark states (from µs to min) which makes difficult dynamic studies using autofluorescent proteins labeling. The photophysical characteristics of these proteins had then to be quantified in the time domain they will be used. More, one had to verify whether their fluorescence is changed from the in vitro to the in vivo case.

In general, use of XFPs at the single molecule level aims at getting access to two important pieces of information : the first one is the localization of the protein with possibly dynamic information. The total number of photons a given molecule emits with no intermittence (no blinking) determines both the precision with which it can be localized[27] and the number of successive images that can be acquired. The best XFPs should be chosen according to this first parameter. The second piece of information comes from the intrinsic 1:1 labeling ratio of the fusion protein with the autofluorescent molecules. A number of proteins function as oligomers. For instance, in neurons, functional metabotropic glutamate receptors are dimers whereas ionotropic glutamate receptors are likely to be tetramers. Precise visualization of the stoichiometry when dynamical studies of neurotransmitter receptors are performed is then a very valuable added value. This can only be performed by good quantification of the fluorescence signal of the individual XFP.

## 2. In-vitro characterization of individual XFPs

The number of photons $N_{det}$, detected from a single molecule in a given experimental arrangement, depends on the collection time $t_c$, the excitation intensity I, the detection efficiency $\eta_{det}$ and the photobleaching/blinking yield at saturation $1/\tau_{bl}^{\infty}$. It is given by:

$$N_{det}(I,t_c) = \eta_{det} k_{\infty} \tau_{bl}^{\infty} \left[ 1 - \exp\left( \frac{-t_c}{\tau_{bl}^{\infty}(1+I_S/I)} \right) \right]$$

where $I_S$ is the saturation intensity, and $k_{\infty}$ the maximum photon emission rate. The exponential term describes the photobleaching and the blinking influence on the signal. In the case when photobleaching and blinking are negligible (i.e. $\tau_{bl}^{\infty}(1+I_S/I) \gg t_c$), the equation converts to the well-known form, $N_{det}(I,t_c) = \eta_{det} k_{\infty} t_c \frac{I/I_S}{1+I/I_S}$

In order to obtain generalized values that are not dependent on the specific experimental parameters, and hence easily comparable between each particular experiment, one can correct the data for the detection efficiency and illumination time effects. Such a generalized quantity is represented by the fluorescence rate:

$$F(I) = \frac{N_{det}}{\eta_{det}\tau_{bl}^{\infty}(1+I_S/I)(1-\exp(-t/(\tau_{bl}^{\infty}(1+I_S/I))))} = k_{\infty} \frac{I/I_S}{1+I/I_S}$$

The mean fluorescence rates (F(I)) of individual eGFP or eYFP have been measured in Harms et al[15] with excitation intensities between 0.5 and 20 kW/cm$^2$. EGFP was excited at 488nm and eYFP at 514 nm. The values for the maximum emission rate, $k_{\infty}$, and the saturation intensity, $I_S$, were then deducted. As a result, when studied in polyacrylamide gels, the mean fluorescence rate of individual eGFP and eYFP are mainly identical ($k_{\infty}$ ~ 3000 photons/ms). Furthermore photobleaching/blinking times are hardly different ($\tau_{bl}^{\infty}$ ~ 3ms) which implies that the maximal number of photons emitted are identical for the two mutants. However, the saturation intensity for YFP is lower ($I_s$ = 7±2 kW/cm$^2$) than for GFP ($I_s$ = 13±3 kW/cm$^2$). This means that less excitation intensity will be needed to detect eYFP than eGFP.

Due to fast photobleaching and poor quantum yield, detection of single fluorescent eCFP is challenging[28] and attempts to fully characterize the fluorescence characteristics of the proteins remained unsuccessful[15].

The autofluorescence present in living cells is the major source of background which may prevent single molecules studies in vivo. Flavinoids which are in abundant concentrations in cells (10$^6$-10$^8$ molecules/cell[29]) are the main source of this cellular autofluorescence in the yellow-green region[30, 31]. The spectral comparison of flavinoid and the various autofluorescence proteins shows that the absorption spectrum of flavinoid strongly overlaps with the excitation spectra of eCFP and eGFP, whereas the emission spectrum overlaps most strongly with that of eYFP, and minor with that of eCFP and eGFP. The saturation intensity of flavins has also been measured in bulk studies, and found to be equal to 35±10 kW/cm$^2$ [15].

Different methods can be used to reduce the cellular background and detect single XFPs. Excitation of the fluorescence through total-internal reflection (TIR) significantly reduces the excited volume which is limited to the evanescent wave region[32, 33]. One disadvantage of TIR is the excitation intensity which decreases exponentially with the distance from the glass. Thus, a precise estimation of the intensity at the molecule position is very difficult given that most cells presents a membrane topology with a variance larger than 150 nm[34]. That might result in more than 95% fluctuation in the excitation intensity of the evanescent field, which renders the analysis of signal amplitude of fluorophores hazardous. A second possibility to reduce background is to perform a short photobleaching treatment with an intense light pulse before the single molecule detection experiment[17]. A third one would be to use confocal scanning microscopy which has the disadvantage to present a slow image acquisition rate in the single molecule configuration. Indeed, the collection time to detect a single XFPs, even at the fluorophore saturation regime can not be arbitrary shortened (typically 1-5 ms is a minimum with a detection efficiency of 5-10%). Parallelism being lost in confocal configuration with respect the use of CCD cameras, acquisition rates of images is limited. A final possibility would be to separated temporally the emission of fluorophores with a long lived excited state from flavins fluorescence by performing a time gated detection technique[35, 36]. Whatever the experimental trick, knowledge of the emission properties of flavins will help to further optimize the experimental conditions. This can be done by optimization of the detection-ratio, R, describing the relative detection yields of the various autofluorescent proteins (XFP) and that of flavinoid (F), $R = \frac{\eta_{XFP}\sigma_{XFP}(\lambda_{XFP})}{\eta_F^{XFP}\sigma_F(\lambda_{XFP})}$ ,[15], for given detection efficiencies, η, and absorption cross sections σ(λ).

One find R = 1.8 for eCFP, R = 8.7 for eGFP, R = 405 for eYFP. Among the different XFPs, the YFP is then confirmed to be far superior to the others for single molecules studies in living cells both from the spectral point of view, and for the emission properties as stated above. It combines a high emission rate and a high detection-ratio with excitation at 514 nm.

## 3. In-vitro characterizations of individual *Ds*Red

While the eGFP and its mutants have proved to be suitable for many applications[13, 37] (see below), the relatively high quantum yield of photobleaching leaves room for improvement in applications which require a large number of emitted photons, such as in single-molecule studies. Also, the most red shifted mutant (eYFP) is not a strong emitter at long (>550 nm) wavelengths, where the interference from flavins is substantially less severe.

Considering these issues, the discovery of a highly fluorescent red protein from the *Discosoma* genus of coral[14], *Ds*Red (originally termed drFP583), has aroused intense interest[38]. The protein is a superior emitter at long wavelengths when compared with GFP and its mutants, with a monomer molar extinction coefficient of 75000 L mol$^{-1}$ cm$^{-1}$ at 558 nm, and a 70% fluorescence quantum yield[39]. For example, the detection-ratio criteria R introduced above gives R>10$^4$ for *Ds*Red. Recent FCS [40] and analytical ultracentrifugation studies [39] have provided strong evidence that the protein is tetrameric even at nanomolar concentrations. With wild-type emission at 583 nm, and the possibility to shift the fluorescence to longer wavelengths through strategic mutagenesis, *Ds*Red may become the cellular reporter protein of choice.

*Ds*red is well excited by the 532 nm laser line of frequency doubled Nd:Yag laser. Harms et al [15] measured the saturation intensity of single *Ds*Red when embedded in polyacrylamide gels, and a higher value is found in comparison to XFPs : (I$_s$ = 50 ± 10 kW/cm$^2$). Furthermore, *Ds*Red is much brighter and more photostable than XFPs[41]. However, the photophysics of *Ds*Red appears to be much different from the XFP's one but not less complicated. Indeed, single-molecule time traces of *Ds*Red reveals multi-step photobleaching behaviors as evidenced by Lounis et al[41]: the intensity dropped either once (~45% of all cases analyzed in [41]), twice (also ~45%), or three and four times in ~10% of all cases. This behavior is conceptually well consistent with the multimeric character of *Ds*Red. More strikingly, the height of the steps as well as their duration is not independent of the number of the step: the majority of the traces that displayed more than one step typically exhibited a large first drop followed by a period of weaker fluorescence before complete photobleaching to the background level. As proposed by Lounis et al, this emissive behavior may result from the coupling between the four chromophores via dipole-dipole energy transfer. In this model, the first period of high brightness corresponds to the fluorescence of an intact oligomer. After sometime, at least one of the constituent fluorophores bleaches under excitation, and the residual fluorescence from the remaining intact chromophores is either negligible or greatly reduced by trapping on the damaged chromophore(s). The calculated rate of energy transfer between intact fluorophores within a complex is significantly faster than the rate of emission[41]. Therefore, if the bleached fluorophore still absorbs efficiently but emits poorly, it can act as a trap for the energy of the absorbed photon and provide a suitable pathway for dissipating the energy via means invisible to our detection methods. Similar quenching effects have been observed in electronically coupled system like single-polymer spectroscopy[42] and photosynthetic antenna complexes[43]. One can also note that epifluorescence ensemble measurements performed on *Ds*Red exhibits two time scales of photobleaching[41] which can also be interpreted by the single molecule observations: while the faster rate may represent the photobleaching of the intact protein, the slower rate may denote the bleaching rate for a chromophore that can dissipate its excited-state energy via nonradiative and non-damaging mechanisms.

## 4. Two-photon excitation

In order to largely increase the signal-to-background ratio for single-molecule detection of autofluorescent proteins in live cells, one could envisage to further shift the wavelength of the excitation light to the red by using two photon excitation. The basic advantage for using two-photon excitation is resting on the earlier observation that the two-photon absorption cross-section of a variety of fluorescent molecules scales super-linearly with the one-photon absorption cross-section[44]. Hence, the ratio of the effective excitation rate of a fluorophore with high one-photon absorption cross-section, like the fluorescent proteins, and a fluorophore with a low one-photon absorption cross-section, like flavines, will be largely increased for two-photon excitation.

The second advantage of two-photon excitation is that it constitutes a good method to scan a three-dimensional tissue space with laser excitation configured to excite two (or more) photons at discrete locations in the tissue[45]. This particular applicability for depth discrimination in thick tissues results from the fact that fluorescence signal intensity is highly dependent on the degree of photon flux, which decreases rapidly as a function of distance from the focal plane.

Measurements of the one and two-photon absorption cross-section of XFPs in bulk[46] show a super-linear scaling behavior which has been previously described only for small organic molecules. Furthermore, the two-photon absorption spectra closely follow those measured for one-photon absorption. However, a significant blue shift is found in the two-photon absorption cross-section of the autofluorescent proteins which can be attributed to the participation of a vibrational mode in the two-photon absorption process. In a similar way to the one photon case, one can generalize the excitation-wavelength dependent relative detection yield for the autofluorescence proteins to the two photon case[46]:

$$R_{OPE/TPE}(\lambda) = \frac{\eta \sigma(\lambda)}{\eta_{flavine} \sigma_{flavine}(\lambda)}$$

where the experimental detection efficiencies for the various fluorophores, $\eta$, and the absorption cross-sections, $\sigma_{OPE}$ and $\sigma_{TPE}$, for the one- or two-photon case are considered. In the case of two-photon excitation $\eta/\eta_{flavine}$ was close to unity. As a result, One find $R = 36$ at 860 nm for eCFP, $R = 467$ at 920 nm for eGFP and $R = 566$ at 960 nm for eYFP[46]. As predicted, the autofluorescence should not be a limitation to the observation of single autofluorescent proteins in living cells using two photon excitation. In fact, the major limitation is due to fast photobleaching of the autofluorescent proteins. In the one photon case it does not prevent to observe single molecules, but in the two photon case it might be a severe limitation. Indeed, it has been found that the process is enhanced in the two photon case with respect to one photon excitation[47].

## 5. Fluorescence Resonance Energy Transfer

For the last three decades, FRET has been extensively used to study intra- and inter-molecular conformational changes as well as to measure distance on the nanometer scale in biological systems. FRET is the nonradiative transfer of electronic excitation energy from donor to acceptor dye molecules by a weak dipole-dipole coupling mechanism. The transfer efficiency was predicted by Förster to decrease with R, the distance between the two dyes, as $1/(1+(R/R_0)^6)$[48]. $R_0$, the Förster radius, is the distance corresponding to 50% energy transfer and depends on the photophysical properties of the dyes and their relative orientations. It allows distance measurements on the 2- to 8 nm scale depending on the FRET pair used and hence is well suited to study conformations of biological macromolecules. At the single-molecule level, FRET has first been observed between two dyes on a dry surface[49] and has been applied to ligand-receptor colocalization on a membrane[50] and to study immobilized proteins and protein-DNA interactions in solution[51]. All those pioneer studies used rhodamine dyes and Cy5 as a FRET pair. Among the different mutants of the GFP and *Ds*Red, 3 pairs can be used to monitor protein-protein interaction via FRET[52]: the preferred one is eCFP/eYFP, whereas, eGFP with *Ds*Red[53] and eBFP (a blue variant of the eGFP excited with UV light) and eGFP can be used. At the single molecule level, the possibility to use single XFPs in FRET studies has been demonstrated in vitro situation using the calcium dependent construct, so-called, cameleon [28, 54]. It contains both one eCFP and one eYFP separated by a calmodulin linker. In a solution calcium free, no FRET is present between the eCFP (the donor) and eYFP (the acceptor) due to a large intra-molecular distance but on calcium binding, the proteins folds in a manner to provoke FRET between the two XFPs. However, up to know, no single molecule experiment has been reported in live cells using the XFPs presumably due to the difficulty to detect single XFPs alone. One can still predict that the better understanding of the photophysical characteristics of the fluorophore and better handling of the cell preparation for single molecule studies will help to go further in that direction. One should also mention that eCFP being difficult to detect, further progress of molecular biology to improve the mutants of the GFP, and the possibility to use *Ds*Red coupled to eGFP for FRET studies[53], will help to perform successful single molecule FRET studies using XFPs in live cells.

## 6. Fluorescence Correlation Spectroscopy

Fluorescence correlation spectroscopy (FCS) analyzes the fluctuations in the fluorescence emission of small molecular ensembles in a very small volume. It was first developed as an alternative way of measuring the translational[55] and rotational[56] diffusion coefficients and chemical reaction rates of molecules in solution. It can provide information about a multitude of parameters, such as concentrations, molecular mobility or dynamics of fluorescently labeled molecules. Performed within diffraction-limited confocal volume elements, FCS is often proposed to provide a way for determining intracellular mobility parameters of very low quantities of fluorophores[57]. It is in fact directly linked to single molecule studies as correlations in the fluorescence signal are best contrasted when the number of fluorophores in the detection volume is in average less or equal to one. However, phenomenon such as photobleaching and intramolecular dynamics, which introduce fluorescence flickering must be carefully controlled when performing FCS studies to avoid unwanted artifacts.

In the general frame of this review, one can ask the question whether combining FCS methods and XFPs would give access to pertinent parameters when used in living cells. The longest correlation time accessible in FCS curves is given by the diffusion time of the molecules through the detection volume providing that the molecule does not photobleach before quitting the detection volume.

For a typical membrane protein, the maximal diffusion coefficient is of the order of $D \sim 0.1 \mu m^2/s$ and for a typical FCS measurement, the surface in which the protein is illuminated is $S \sim 0.1 \mu m^2$. The time it takes for the molecule to diffuse through the surface is then given by $S/4D \sim 250ms$. Fig.2 gives the mean photobleaching time measured for eYFP molecules immobilized in polyacrylamide gels for two different illumination intensities. Photobleaching curves were fitted by : $G(t) = N^{-1} \times \exp(-t/t_f)$. One find $t_f$=10 ms for 4KW/cm$^2$ and 250 ms for 0.6KW/cm$^2$. For slowly diffusing proteins, very low illumination intensities should then be used as in[17]. Quantitative analysis of slowly diffusing proteins is then difficult due to poor signal-to-noise[58]. It then becomes very difficult to perform accurate measurements on XFPs using FCS alone. FCS could then be more suitable to very fast processes such as fast diffusing

XFPs molecules[58] or molecular interactions, providing that the internal photophysics of the chromophores is fully understood. Indeed XFPs themselves produces a high number of dark states present at almost all time scales (microseconds to seconds [25, 26]). These dark states could then be difficult to distinguish from the physical parameters under study.

## 7. Live cells studies

Among the different autofluorescent proteins, eYFP appears to be the best choice to perform single molecule studies in live cells, due to its brightness, spectral separation to flavins and absence of oligomerization. Studies in vitro are however not sufficient to ensure that single eYFP will be detectable in vivo. It is necessary to compare the signal of individual eYFP when fused to a protein expressed in a cell. Fusion of eYFP to pore-forming $\alpha_{1C}$-subunit of cardiac L-type $Ca^{2+}$ channel (eYFP-$\alpha_{1C}$) in HEK293 cells were studied[17] as well as to metabotropic glutamate receptors for neurotransmettor (mGluR5) in COS-7 cells at the single molecule level (see fig.3). Single eYFP are detectable with a very good signal to noise ratio (>10 on Fig.3). Furthermore, we found no noticeable difference in the fluorescence characteristics of single eYFP detected in vitro and in live cells. This indicates that in vitro calibration of the fluorescence eYFPs can be used for in vivo measurements.

Due to the previously outlined difficulties to detect single XFP in cells, few articles relate the study of single XFP-fusion proteins in cells[15-17]. Following the first report of the observation of single autofluorescent protein in a living cell (eYFP-$Ni^{2+}$:NTA:DOGS lipids in the plasma membrane of HASM cells) by Harms et al, Iino et al used eGFP to reveal E-cadherin oligomers in living mouse fibroblast L cells. The choice of the eGFP mutant forced Kusumi's group to excite the fluorophore in total internal reflection. As pointed before, this excitation method, although efficient to remove cellular background might lead to misinterpretation in analyses of signal amplitude. Furthermore, internalization of the membrane proteins by endocytosis can not be distinguished from photobleaching in such a geometry. Stoechiometry is better accessible using wide field excitation which also permit to study proteins in other places than the basal membrane[17]. Indeed, studies of eYFP-$\alpha_{1C}$ fusion proteins at the single molecule level in HEK293 cells using wide field imaging has permitted to measure the degree of aggregation of the $Ca^{2+}$ channel subunits[17]. Comparison of the intensity distributions of the signals detected in the live cells and that obtained for purified eYFP show that while the intensity distribution obtained for purified eYFP is unimodal (close to a Gaussian curve), that for eYFP-$\alpha_{1C}$ expressed in HEK293 cells is multimodal. As a consequence, analysis of the distribution permits to calculate the mean number of proteins in a cluster which is found to be on the order of 40[17].
The analysis of the diffusion of single biomolecules via autofluorescent proteins even if possible using eYFP [17] is strongly limited to very short trajectories due to photobleaching as described above. We recently tested citrine, a new variant of the eYFP[40, 59], which has been reported to be twice as more stable than eYFP in bulk studies[59]. Preliminary results at the single molecule level using eYFP-mGluR5 or Citrine-mGluR5 constructs expressed in COS-7 and PTK2 cells show that the use of citrine could significantly increase the stability of the fluorophores. As an example, Fig.4 show a typical trajectory of 10 points obtained from a single Citrine-mGluR5 in the plasma membrane of a COS-7 cell. In the same condition (illumination intensity of 4kW/cm$^2$ at 514 nm during 3ms/point), the typical trajectory length obtained from eYFP would be 3 to 5 points. Complete characterization of the citrine with respect to eYFP is in progress at the single molecule level.

## 8. Conclusion

This review proposed a state-of-art of the possibility to detect and image single autofluorescent proteins for studies in living cells. The characterization of different mutants of the GFP and comparison to *Ds*Red in various biocompatible in vitro environments permits to conclude that to date eYFP (or even one of its mutant, citrine) is likely a superior choice for applications in dynamics of individually labeled fluorescent fusion proteins. This is due to its brightness, resistance to photobleaching (especially in the case of citrine), and detection ratio. However, there is still room for improvement as the origin of blinking is not completely elucidated and efforts are being made to optimize the fluorescence of these proteins. The possible advantages of *Ds*Red, which are its long-wavelength emission, its resistance to photobleaching, and high fluorescence quantum yield, are good stimulants for further molecular biology improvement of the chromophore. It suffers still from its tetramerization behavior and multicolor emission[60, 61].

The viability of single-molecule fluorescence in live cell studies was also presented through different examples. In particular, the possibility to "count" the molecules in a cell is a very promising feature of the use of the autofluorescent proteins in biology. One should mention still that in live cells, experiments with individual autofluorescent proteins are a challenging task and that further research and technological advancements will be needed before the methods is used as a routine in biology laboratories.

Acknowledgments: We wish to warmly thank Thomas Schmidt for helpful discussions and Jacqueline Butter for taking the data on citrine. We thank Françoise Rossignol for help in performing the citrine constructs. This work was supported by grants from the CNRS, the Conseil Régional d'Aquitaine and the Ministère de la Recherche.

Figure 1:

Diagram of a typical wide-field single-molecule fluorescence microscope.

Figure 2:

Single-molecule fluorescence detection of mGluR5-eYFP in the plasma membrane of COS-7 cells. (**A, B**) Images of the same area as seen by white light imaging (**A**) and epifluorescence (**B**). (**C**) 3D representations of the same molecules. (**D**) Fluorescence intensity versus time trace of a single eYFP-mGluR5 molecule showing one step photobleaching. The sample was excited with 514 nm laser light for 5 ms/point at 4 kW/cm$^2$.

Figure 3:

Photobleaching of eYFP immobilized in polyacrylamide gels evidenced by Fluorescence Correlation Spectroscopy for two different excitation intensities.

Figure 4:

Single-molecule fluorescence detection of mGluR5-citrines in the membrane of a PTK2 cells. The samples were excited with 514 nm laser light for 3 ms/point at 4 kW/cm$^2$. (**A**) 3D representation of the fluorescence signal for a single mGluR5-citrine. (**B**) Fluorescence intensity versus time of the same molecule showing both blinking and one step photobleaching. (**C**) Trajectory reconstruction of the same molecule.

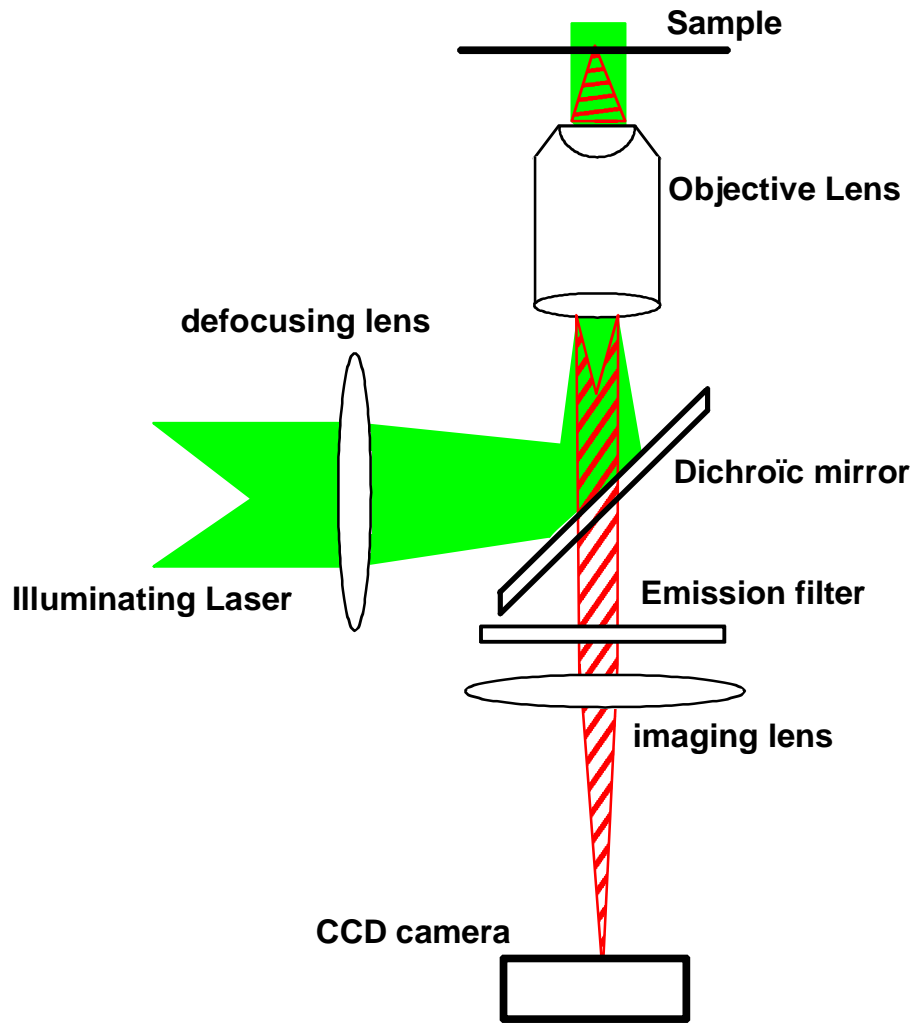

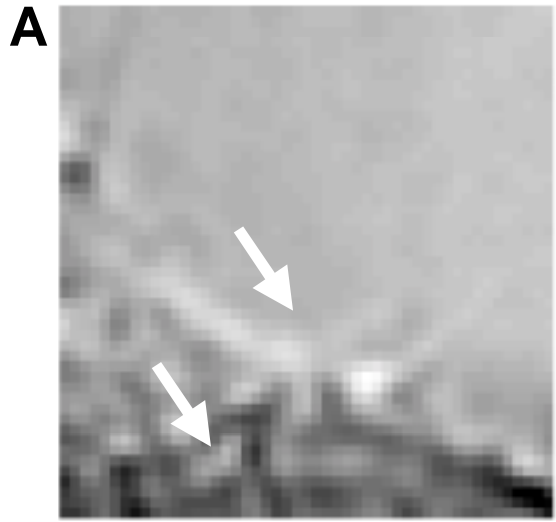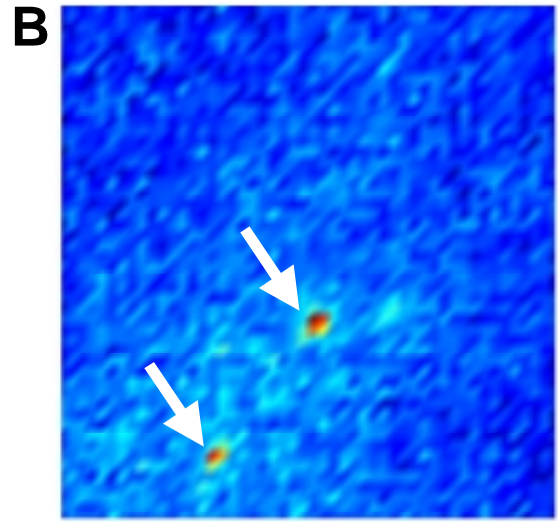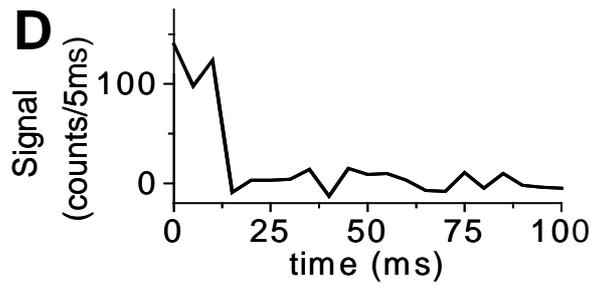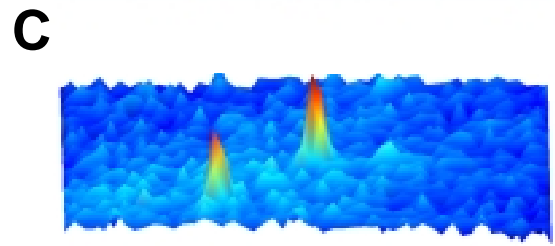

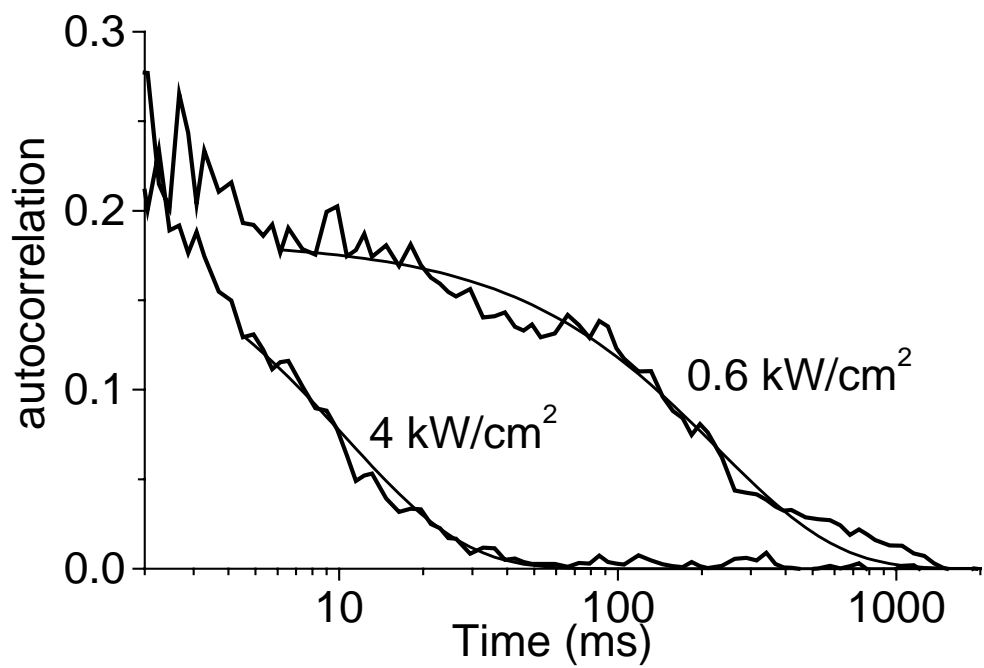

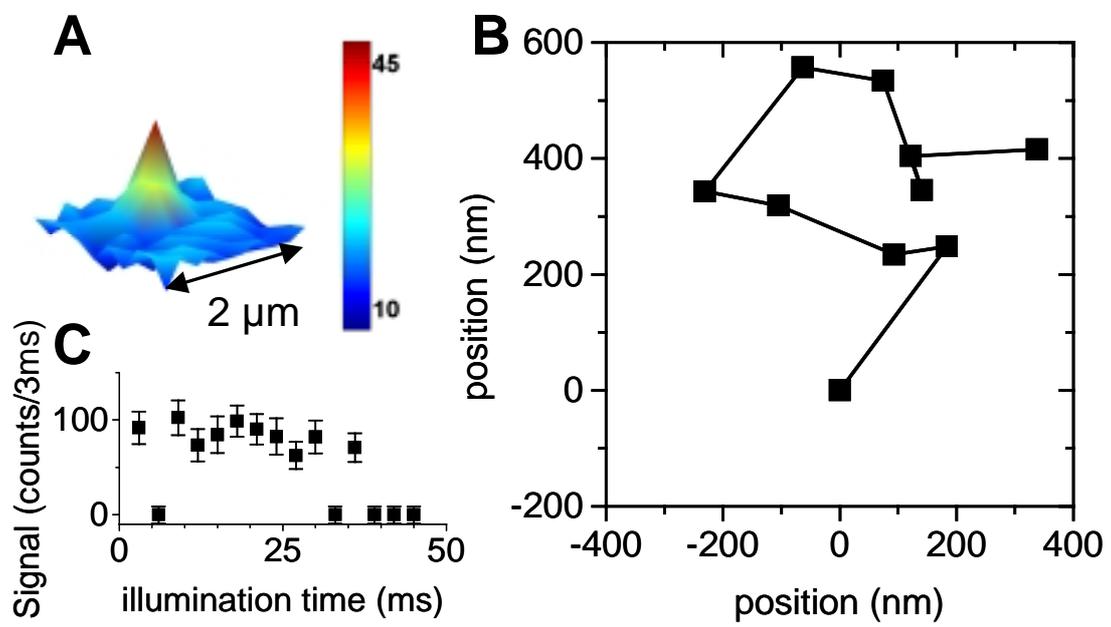